\documentclass[aps,prb,twocolumn]{revtex4-1}
\usepackage{graphicx}
\usepackage{times}
\usepackage{color}
\input pdfcolor.tex
\usepackage{graphicx}
\usepackage{epsfig,color}
\begin{document}

\title{
Population imbalanced lattice fermions near the BCS-BEC crossover: \\
II.~The FFLO regime and its thermal signatures}

\author{Madhuparna Karmakar and Pinaki Majumdar}
\affiliation{Harish-Chandra Research Institute, 
Chhatnag Road, Jhunsi, Allahabad 211 019, India}

\begin{abstract} 
We study the effect of high population imbalance in the two dimensional 
attractive Hubbard model, in the coupling regime corresponding to BCS-BEC 
crossover, and focus on thermal effects on the Fulde-Ferrell-Larkin-Ovchinnikov 
(FFLO) state. Using a method that retains all the classical thermal fluctuations
on the FFLO state we estimate a very low $T_c$ and infer a strongly first order
normal state to FFLO transition. The $T_c$ is an order of magnitude below the 
mean field estimate. We track the fermionic momentum distribution, the density 
of states, and the pairing structure factor deep into the normal state. The 
pairing structure factor retains weak signature of finite momentum pairing  
to a high temperature despite the low $T_c$ itself, while the spin resolved 
density of states changes from the `pseudogapped' FFLO character to gapless and 
pseudogapped again with increasing temperature. These results map out the rather 
narrow temperature window, and diverse physical indicators, relevant to FFLO states 
on a cold atom optical lattice, and complements our work on the lower field 
`breached pair' state.
\end{abstract}

\date{\today}

\maketitle

\section{Introduction}
The last decade has seen an intense search for spatially modulated 
superconducting (or superfluid) states which were originally 
predicted by 
Fulde and Ferrell (FF) \cite{ff} and Larkin and Ovchinnikov (LO) 
\cite{lo}. These modulated paired states are expected to arise in
the presence of a strong magnetic field that Zeeman couples to the
fermions, but in most traditional superconductors the
`orbital' suppression, via Abrikosov lattice formation,
takes place before any spin related effects can show up.
This led to a decades long wait before a heavy fermion,  
\cite{bianchi2003,kenzelmann2014,tayama2002,mitrovic2008,kumagai2006,mitrovic2006,wright2011,
kenzelmann2008}, CeCoIn$_5$,
some quasi-2D organic superconductors of the $\kappa$-BEDT family
\cite{lortz2007,beyer2013,coniglio2011,mayaffre2014,bergk2011,agosta2012,cho2009}, 
and the ferropnictide \cite{zocco2014,prozorov2011,kim2011,ptok2013,ptok2014} 
KFe$_2$As$_2$ 
were identified as promising `solid state' candidates.
Developments in cold atom physics, on the other hand,
allowed the exploration of pairing among population imbalanced
neutral fermions, without the complication of orbital (Lorentz
force) effects \cite{partridge2006,shin2008,ketterle_science2007,shin2006,liao2010,ketterle_science2008}. 

The modulated, or finite momentum,
character of the paired state
is difficult to probe since external
fields do not couple directly to the pairing order parameter.
Indirect probes in the solid state, for example, have probed 
(i)~thermodynamic features like 
specific heat and magnetization \cite{bianchi2003, lortz2007, bergk2011, beyer2013},  
(ii)~spectroscopic aspects like NMR shift and relaxation \cite{kumagai2006,
mitrovic2006, mayaffre2014, wright2011},
or (iii)~spatial modulation of the induced 
magnetization, and have yielded 
suggestive results for the high field state in CeCoIn$_5$
\cite{bianchi2003,kenzelmann2014,tayama2002,mitrovic2008,kumagai2006,mitrovic2006,wright2011,
kenzelmann2008},
the organics \cite{lortz2007,beyer2013,coniglio2011,mayaffre2014,bergk2011,agosta2012,cho2009},
and the pnictides \cite{zocco2014,prozorov2011,kim2011,ptok2013,ptok2014}.
In cold atomic gases the direct spatial signature
of a modulated superfluid state continues to 
be elusive but the density profiles of the up and
down spin condensates 
give clear indication of population imbalance
\cite{partridge2006,shin2008,ketterle_science2007,shin2006,liao2010,ketterle_science2008}. 
The finite spin polarization indicates 
\cite{ketterle_science2008,liao2010,shin2006}
the coexistence of unpaired fermions with the superfluid 
condensate down to zero temperature.  

There is a large body of theoretical work exploring
modulated mean field states in both the continuum
and lattice cases. The mean field theory (MFT) 
is non trivial since the nature of modulation,
{\it i.e}, the wave vectors involved, is sensitive to
the applied field or population imbalance \cite{trivedi,torma2007}. Nevertheless,
the density and field (or magnetization) 
window over which a FFLO ground state is expected is
reasonably mapped out 
\cite{trivedi, torma2007, scaletter2012, scaletter2008, casula2008,zhang2013} 
in lattice models. In fact in two dimensions there now
exists a detailed characterization of the FFLO ground
state based on extensive variational calculations \cite{zhang2013}.

What most of these calculations do not address is the
thermal window over which the FFLO state is expected to
survive. 
Mean field theory has been extended to finite 
temperature \cite{trivedi,torma2007} 
but in the BCS to BEC crossover regime, where
one expects the $T_c$ to be highest,  the results 
are very unreliable. In three dimensions, for couplings
chosen to obtain the peak zero field 
$T_c$ in the BCS-BEC crossover
window, the FFLO
$T_c$ is estimated to be about 
half the zero field $T_c$, which mean field
theory sets as $\sim 1.1t$ (where $t$ is the 
hopping amplitude).
This is a severe overestimate \cite{beck}!
Dynamical mean field theory 
\cite{torma2012, torma2013, torma2014} 
(DMFT) has been employed, both in the `real space' as well as the
cluster form, but typically addressing only strongly anisotropic
lattices.
Finally, there is quantum Monte Carlo
(QMC) data in two dimensions \cite{scaletter2012}
which sets the zero field $T_c$ as $\sim 0.1t$ (and the
FFLO $T_c$ obviously less than that) but suggests that
FFLO correlations survive to a scale $T \gg T_c$.

As the numbers above indicate, estimates of the thermal stability
window vary over a wide range.
The existing results also do not 
address the {\it spectral features} of the thermal 
FFLO state, or
the normal state with FFLO correlations. We feel there is
room for a less computation intensive, possibly more
intuitive, approximation that retains the rich detail 
of the mean field ground state but also accesses the
key thermal fluctuations. 

We address the thermal physics in the FFLO window, 
with the coupling set to the BCS to BEC crossover regime,
in the two dimensional attractive Hubbard model.
Using an auxiliary field based real space Monte Carlo
we observe the following:
(i)~The maximum $T_{c}$ for FFLO phases
is about $1/6$ of the zero field $T_c$, and about 20\% of
what mean field theory predicts.
In absolute terms it is just about 
two percent of the hopping scale.
(ii)~The FFLO to normal transition is strongly first order
and the pairing structure factor retains a weak
signature of finite momentum pairing to several times the
$T_c$ scale.
(iii)~The spin resolved 
density of states evolve from a `pseudogapped' form,
characteristic of LO order, at
low temperature to a gapless form above $T_c$, but then
loses weight at the Fermi level as the temperature is
further increased. 

The paper is organized as follows. In 
Section II we briefly touch upon our
model and method since most of this is 
discussed in detail in Paper I \cite{mpk_bp}.
 Section III presents our results on 
the variational ground state, and 
Monte Carlo inferred thermodynamic, 
spatial, and spectral features.
Section IV discusses some issues and
benchmarks related to our 
numerical method.
We conclude in Section V.
Much of the background material and the low field
results are discussed in our Paper~I.

\section{Model and method}

\subsection{Model}

We study the attractive two dimensional Hubbard model 
(A2DHM) on a square lattice
in the presence of a Zeeman field:
\begin{equation}
H  = H_0  
- h \sum_{i} \sigma_{iz}  
- \vert U \vert \sum_{i}n_{i\uparrow}n_{i\downarrow} \label{eq1}
\end{equation}
with, 
$ H_0 =  \sum_{ij, \sigma}(t_{ij} - \mu \delta_{ij}) 
c_{i\sigma}^{\dagger}c_{j\sigma}$,
where 
$t_{ij} = -t$ only for nearest neighbor hopping and is zero otherwise.
$ \sigma_{iz} = (1/2)(n_{i \uparrow} - n_{i \downarrow})$.
We will set $t=1$ as the reference energy scale.
$\mu$ is the chemical potential and
$h$ is the applied magnetic field in the ${\hat z}$ direction.
$U > 0$ is the strength of on site attraction. 
We will use $U/t=4$.

\subsection{Methods}

Methodological issues have been discussed in detail in 
earlier papers \cite{evenson,tarat2014,mpk_bp} 
so we  put in only a brief description for
completeness.
The interaction is decoupled using
a space-time varying 
complex auxiliary field $\Delta_i(\tau)$
 in the pairing channel and we retain 
only the zero frequency mode of the 
auxiliary field, {\it i.e} make a 
static auxiliary field (SAF)
approximation. However,
the spatial fluctuations of $\Delta_i$ are
completely retained. 
The SAF scheme leads to the effective Hamiltonian:
\begin{equation}
H_{eff}  =  H_0 - h \sum_{i} \sigma_{iz} 
+ \sum_{i}(\Delta_{i}c_{i\uparrow}^{\dagger}c_{i\downarrow}^{\dagger}
+ h.c) 
+ H_{cl}  \label{eq2}
\end{equation}
where $H_{cl} = 
 \sum_{i}\frac{\mid \Delta_{i} \mid^{2}}{U} $
is the stiffness cost associated with the auxiliary field \cite{mpk_bp}.
We use two strategies to study the model above:

I.~{\it Monte Carlo:} We generate the equilibrium $\{\Delta_{i}\}$ 
configurations by iteratively diagonalising
the electron Hamiltonian $H_{eff}$ for 
every attempted update of the 
auxiliary fields. 
The Monte Carlo is implemented using
a cluster approximation, in 
which instead of diagonalising the entire
$ L \times L$ lattice 
for each local update of the $\Delta_i$  a 
smaller cluster, of size $L_c \times L_c$,
 surrounding the update site is diagonalised 
 \cite{tarat2014,mpk_bp}. 

II.~{\it  Variational calculation:} 
The zero temperature limit within the SAF scheme is
equivalent to unrestricted 
minimization of the ground state energy over 
configurations of the field $\Delta_i$.
We have carried out minimization of the
energy at  several values of $\mu$ and $h$,
exploring the following kind of periodic configurations:
(i)~`axial stripes':  
$\Delta_{i} \sim \Delta_0 \cos(qx_i)$, 
diagonal stripes 
$\Delta_{i} \sim \Delta_0 \cos(q(x_i + y_i))$,   
(ii)~two dimensional 
modulations, $\Delta_{i} \sim \Delta_0 (\cos(qx_{i})+ \cos(qy_{i}))$, 
and of course (iii)~the unpolarised
superfluid (USF) state $\Delta_i = \Delta_0$.
We minimize the energy with 
respect to the $q$, and $\Delta_0$.

III.~{\it Green's functions for $T=0$:}
It is useful to set up a {\it low order approximation}
 for 
the Green's function of the electron, applicable in
the $T=0$ variational state, for $\Delta_{0}
<< zt$, where the coordination number $z=4$ in 2D. 
In FFLO the $\Delta_{0}$ is strongly suppressed due to 
the magnetic field and the electron pairing takes 
place between the ${\bf k}$ and $-{\bf k} \pm {\bf Q}$ states. 
The resulting spin resolved Green's functions
take the approximate form:
\begin{eqnarray}
 G_{\uparrow \uparrow}({\bf k}, i\omega_{n}) & = & \frac{1}
 {i\omega_{n} - (\epsilon({\bf k})-\mu_{\uparrow}) - 
 \Sigma_{\uparrow \uparrow}({\bf k}, i\omega_{n})} \cr
 \Sigma_{\uparrow \uparrow}({\bf k}, i\omega_{n}) & = & 
 \frac{\Delta_{0}^{2}}{4}[\frac{1}{(i\omega_{n} + 
\epsilon({\bf -k-Q})-\mu_{\downarrow})} \cr
 &&~~~~~~~ + \frac{1}{(i\omega_{n}
+\epsilon({\bf -k+Q})-\mu_{\downarrow})}] \nonumber 
 \end{eqnarray}
where, $\epsilon({\bf k}) = -2t(\cos(k_{x})+\cos(k_{y}))$. 
A similar expression can be obtained 
for $G_{\downarrow \downarrow}({\bf k}, i\omega_{n})$
as well.

One can extract the spectral function $A_{\uparrow \uparrow}({\bf k}, \omega)
= -(1/\pi)Im G_{\uparrow \uparrow}({\bf k}, 
\omega + i\eta)\mid_{\eta \rightarrow 0}$
from the above equation, whose k-sum would give the expression for the 
DOS.

\subsection{Parameter regime and indicators}

The results in this paper are 
at $U=4t$, both within Monte
Carlo and the variation scheme. We have also
explored $U=2t$ variationally and 
observed that the FFLO regime is significantly 
shrunk for a lower value of $U/t$. In fact, in order to 
access the FFLO regime at $U=2t$ we had to use
a larger system size of $L = 36$ since at $L = 24$ (at which
$U=4t$ calculations are being carried out)
no FFLO 
signatures could be observed. For the ground state,
determined variationally, we have checked that the 
character of the $\mu-h$ phase diagram does not 
vary much between $L = 24$ and $L = 48$.
At $U = 4t$ we have explored the $h-T$ dependence at
multiple values of $\mu$ below half-filling (the physics
above half-filling can be inferred from this)
but the qualitative physics
seems similar, so this paper focuses on detailed
indicators at a single $\mu$.
The density at this point is
$n \sim 0.94$, and does not 
vary significantly for the $h$ or $T$ that we have chosen.
We discuss temperature dependence in the 
FFLO regime:  $h/t \sim [0.85:1.25]$.  

As discussed in Paper~I, the phases 
can be classified 
into (a)~unpolarized superfluid (USF), 
(b)~breached pair (BP),
(c)~modulated superfluid (FFLO),
and (d)~a partially
polarized Fermi liquid (PPFL).

We use the following indicators to characterize the physics on the 
$L \times L$ system:
(i)~The structure factors, $S_{\Delta}({\bf q})$ and
 $S_m({\bf q})$ defined earlier \cite{mpk_bp}, 
and $\Gamma({\bf q}) = \sum_{ij} \Gamma_{ij} e^{i {\bf q}.({\bf r}_i
- {\bf r}_j)}$, where 
$ \Gamma_{ij}  =  
\langle c^{\dagger}_{i\uparrow} c^{\dagger}_{i \downarrow} \rangle
\langle c_{j\downarrow} c_{j \uparrow} \rangle $.
(ii)~The bulk magnetization, and the pairing order parameter
$S_{\Delta}({\bf Q})$, where ${\bf Q}$ is the ordering wave vector. 
(iii)~Monte Carlo snapshots of 
(a)~the magnitude $\vert \Delta_i \vert$ of the 
pairing field, (b)~the
correlation $cos(\theta_0 - \theta_i)$ where $\theta_i$ is
the phase of $\Delta_i$ and $\theta_0$ is the phase 
at a fixed reference site on the lattice, 
(c)~the magnetization variable 
$m_i = \langle n_{i\uparrow} - n_{i \downarrow} \rangle $, and
(d)~particle number 
$n_i = \langle n_{i\uparrow} + n_{i \downarrow} \rangle $.
These explicitly highlight the 
modulated nature in the FFLO state and spatial fluctuations with
increasing temperature.
(iv)~The momentum occupation
number $\langle \langle n_{{\bf k} \sigma} \rangle \rangle$
that carries the signature of population imbalance and 
finite ${\bf Q}$ pairing. Finally, (v)~The fermionic 
density of states (DOS) \cite{mpk_bp}. 

\begin{figure}[t]
 \centerline{
 \includegraphics[height=5.3cm,width=4.5cm,angle=0]{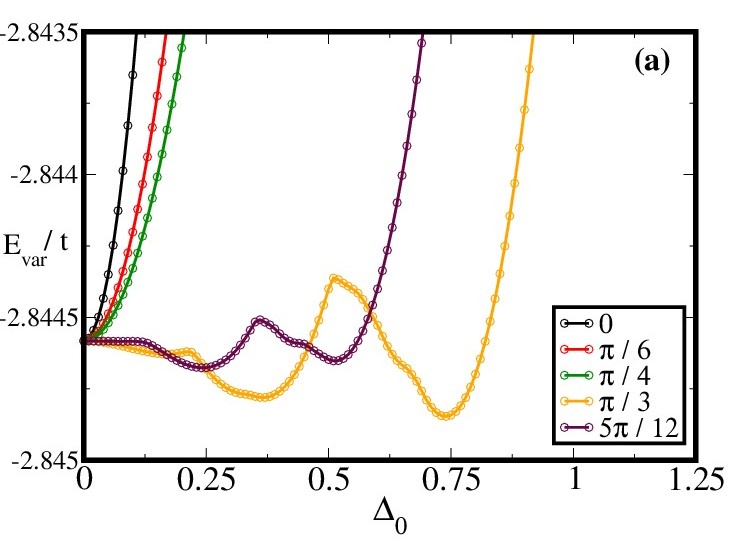}
 \hspace{-0.1cm}
 \includegraphics[height=5.3cm,width=4.5cm,angle=0]{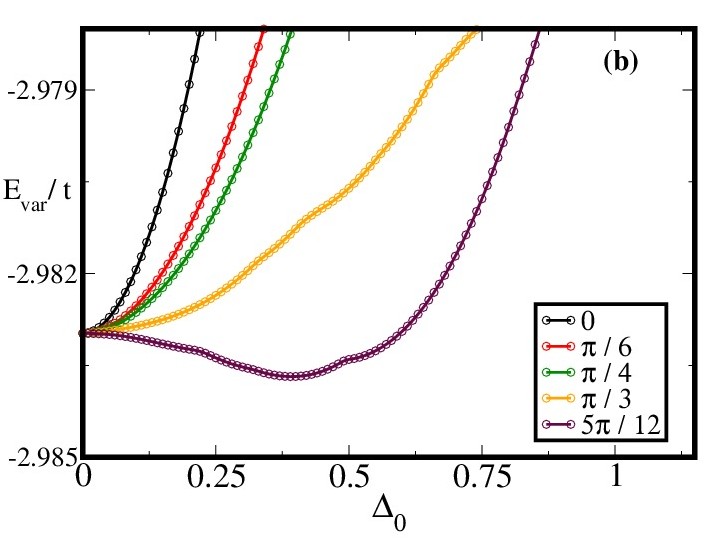}
 }
\caption{Color online: Comparison of energy as obtained
for different axial stripe phases at (a) $h/t = 1.00$ and 
(b) $h/t = 1.10$.
$\Delta_0$ is the strength of the modulation while the 
modulation vectors are ${\bf q} = (q,0)$ with 
$q = 2n\pi/L$. We have compared these also with diagonal stripes and
2D patterns, those data are not shown here.
}
\end{figure}

\begin{figure}[b]
 \includegraphics[height=5cm,width=5cm,angle=0]{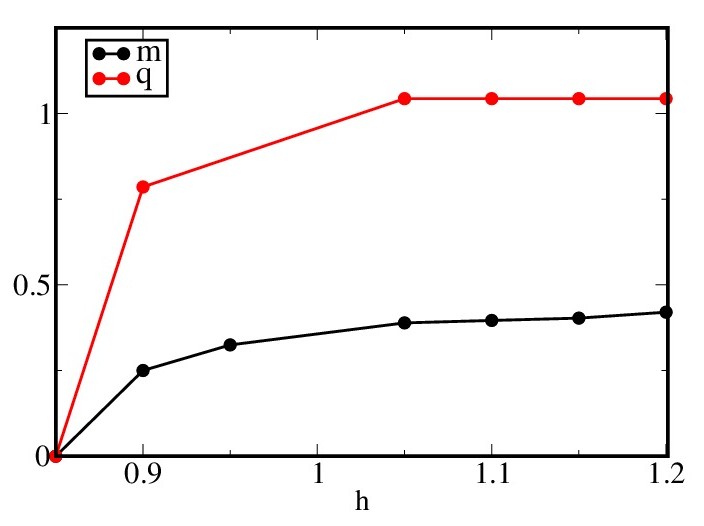}
 \caption{Color online: Field dependence of magnetization $(m)$ and
axial  stripe wave vector $(q)$ at $\mu = -0.2t$ for varying $h$
as obtained from the variational calculation. On our system size we
observe $q \sim 3m$, while a more elaborate analysis suggests
$q = \pi m$.
}
\end{figure}
\begin{figure}[t]
\centerline{
\includegraphics[height=5.0cm,width=4.5cm,angle=0]{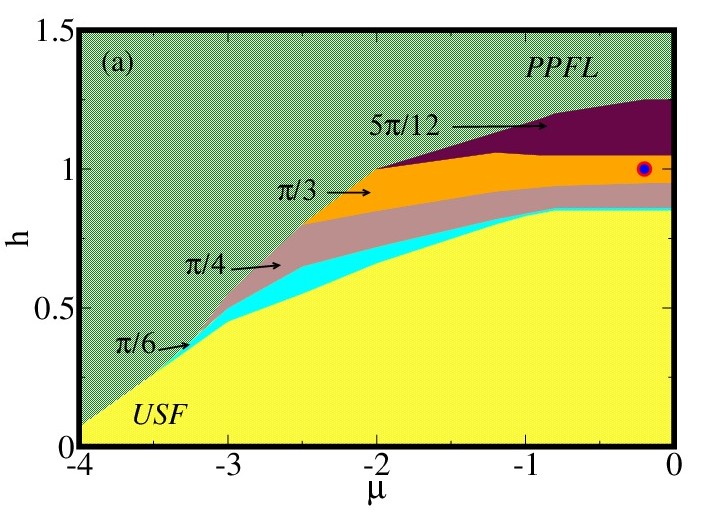}
\hspace{-0.3cm}
\includegraphics[height=4.9cm,width=4.5cm,angle=0]{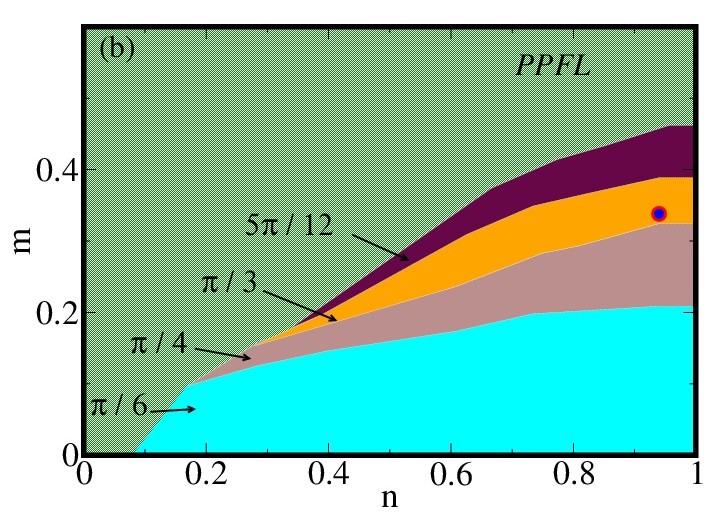}
}
\caption{Color online: Variational ground state in
terms of (a)~varying $\mu - h$ and (b)~varying $n-m$,
showing the different LO phases. The results are on a $24 \times 24$
lattice. The allowed $q$ values are discrete,  $n\pi/L$,
and the transitions between LO phases have small
density discontinuities on this lattice size.
The variation in $q$ is 
expected to be continuous with $m$ as $L \rightarrow \infty$, 
and the density discontinuities would vanish.
The `operating point' for which most of our thermal data 
is shown in this paper is marked by a circle in both panels.
}
\end{figure}

\section{Results: the ground state}

\subsection{Variational phase diagram}

We computed the energy of the SAF Hamiltonian 
for three families of 
trial configurations
$\{\Delta_{i}\}$. These correspond to 
axial stripes, checkerboard modulation,
 and diagonal stripes, 
with different 
wave vectors ${\bf q}$,
as detailed before.
For a $24 \times 24$ 
system we find that the axial
stripes have the lowest energy.
We have checked our results at a larger 
system size of $48 \times 48$ and have found
that the axial stripe phases still turn out 
to be energetically favorable compared  
to the other variational candidates.
A similar calculation by
Chiesa {\it et al.} 
\cite{zhang2013} 
observed that at $U=4t$ 
diagonal and axial stripes make up the ground state
in the relevant $n-m$ window. The possible reason 
for the differences between our results and those
presented by Chiesa {\it et al.} are discussed
at the end of the paper.

Fig.1 shows illustrative data on the 
energy of variational axial stripes,
$\Delta_i = \Delta_0 cos(qx_i)$,
with respect to the amplitude 
$\Delta_0$ for a given $q$. 
The full variational comparison involves
diagonal stripes and checkerboard patterns as well,
we have not shown them for clarity.
Panel (a) shows that the ${\bf q} =0$ USF state
is clearly unfavored and the global minimum is obtained for
${\bf q} = \{\pi/3, 0\}$.
Panel (b), which is at a higher magnetic field, prefers a larger
wave vector, ${\bf q} = \{5\pi/12, 0\}$.

{The energy was calculated deliberately for finite size
systems, $L=24$ in this case, since the Monte Carlo is
done on that size.
The absolute minimum, in the space of $q$ and $\Delta_0$,
 defines the mean field ground state for a given $\mu$
and $h$. The local minima are metastable within the
mean field scheme.} 

{Given the modulated nature of the candidate states
there is a size dependence to some of the features.
By varying $L$ from $16$ to $60$ we
found that (i)~the  minimum ${\bar q}(\mu,h)$
and ${\bar \Delta_0}(\mu,h)$ are reasonably size
independent for $L \ge 20$, but (ii)~the
detailed `energy landscape', as in Fig.1,
 does depend on system size,
and for $L \ge 48$ the local minima vanishes,
leaving us only with the global minimum.
}

\begin{figure*}[t]
\centerline{
\includegraphics[height=5.5cm,width=6.2cm,angle=0]{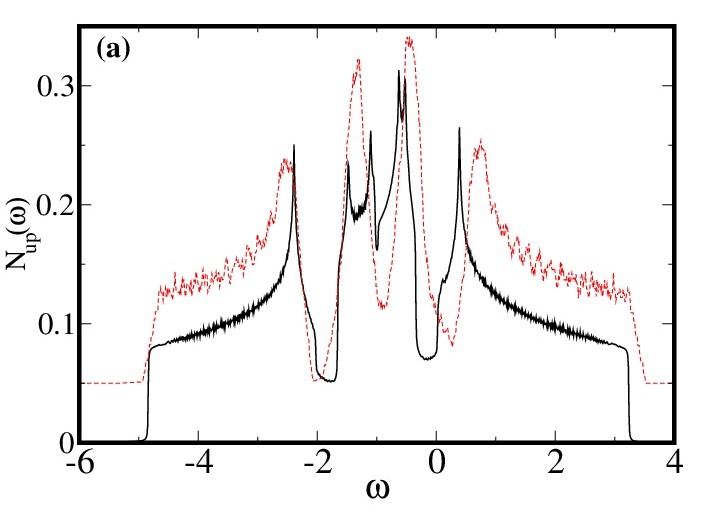}
\hspace{-0.1cm}
\includegraphics[height=5.5cm,width=6.2cm,angle=0]{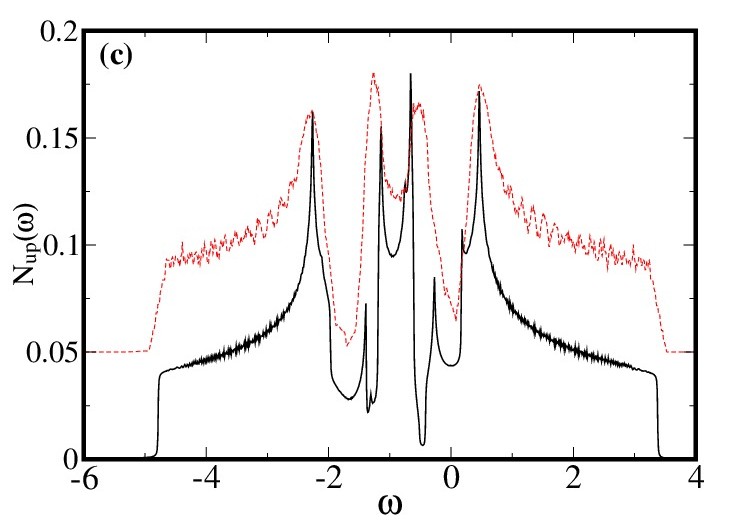}
\hspace{-0.1cm}
\includegraphics[height=5.5cm,width=5.5cm,angle=0]{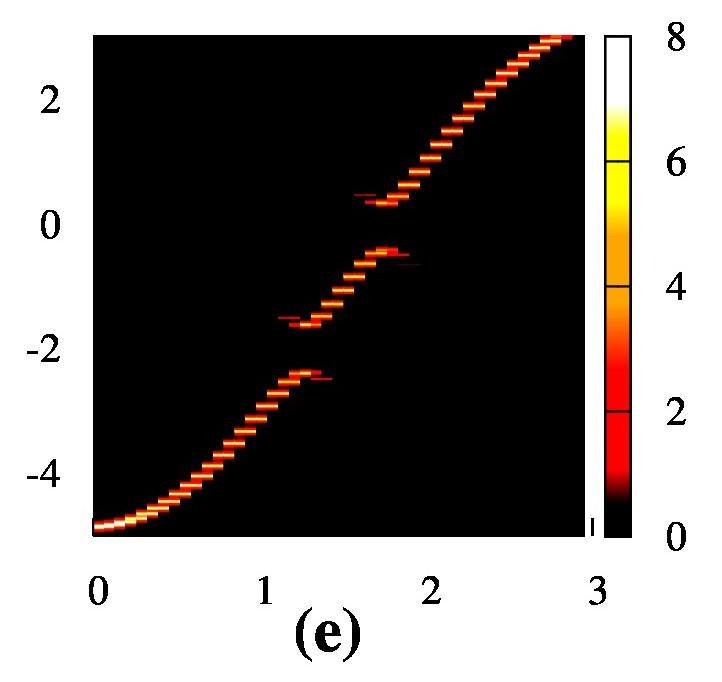}
 }
 \vspace{-0.1cm}
\centerline{
\includegraphics[height=5.5cm,width=6.2cm,angle=0]{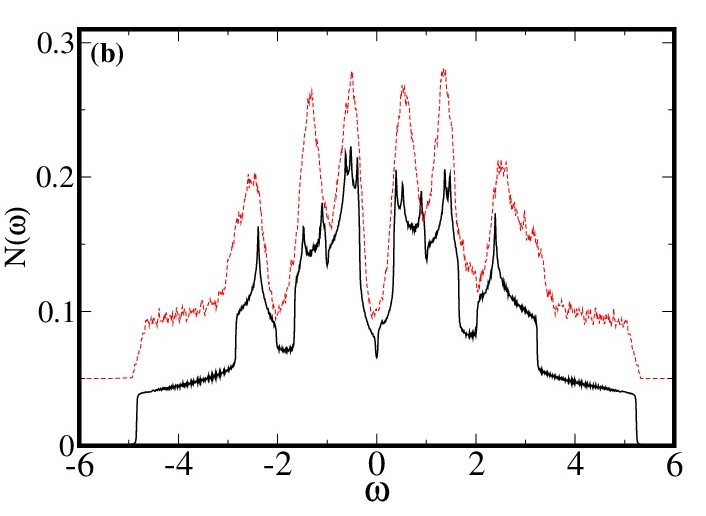}
\hspace{-0.1cm}
\includegraphics[height=5.5cm,width=6.2cm,angle=0]{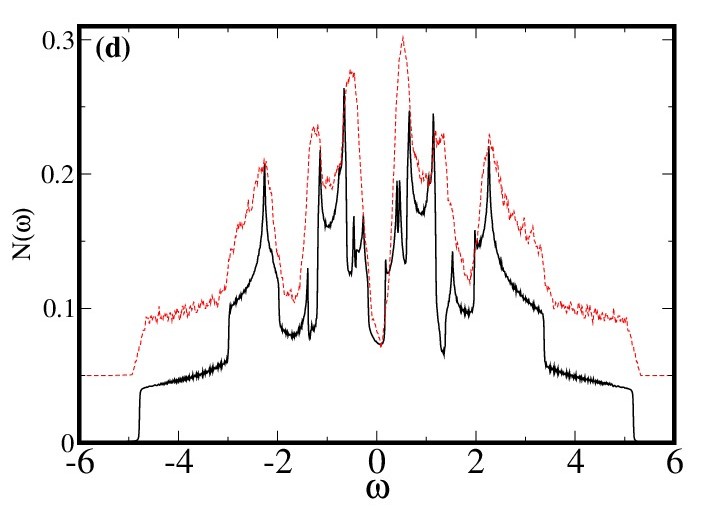}
\hspace{-0.1cm}
\includegraphics[height=5.5cm,width=5.5cm,angle=0]{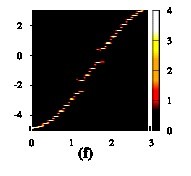}
}

\caption{Color online: Spin resolved and total density
of states for two different modulation wave vectors
calculated through the BdG (red dotted line) and the
Green's function (black solid line) formalism.
The dotted curves are shifted along the y-axis for the
sake of clarity.
(a) and (c) correspond to the spin up DOS for
${\bf q} = \{\pi/3, 0\}$ and $\{\pi/4, 0\}$ respectively,
while (b) and (d) correspond to the total DOS for the same
modulation wave vectors. (e) and (f) shows the
spin up spectral function $A_{\uparrow \uparrow}({\bf k}, \omega)$
calculated for the $\{0, 0\}$ to $\{\pi, \pi\}$ momentum scan
through the Green's function and BdG formalism, respectively,
for ${\bf q} = \{\pi/3, 0\}$.
}
\end{figure*}

{The overall reliability of our phase diagram, even on
$L=24$, is borne out by our observation 
$q \sim \pi m$, Fig.2,   
which was obtained elsewhere
on an {\it infinite system} using a more elaborate mean field
decomposition \cite{zhang2013}.
We show $m$ and $q$ for varying $h$ in
Fig.2. The regime of validity of the relation $q \sim \pi m$ has 
been numerically established \cite{zhang2013} and its basis 
has been analytically discussed \cite{zhang2011} in the literature, 
in the context of half-filling repulsive Hubbard model. The 
repulsive Hubbard model is related to the attractive Hubbard model 
via a particle-hole transformation.

{The $q \sim \pi m$ relation is based on two inputs: 
(i) the knowledge that 
the repulsive half-filling problem has $(\pi, \pi)$ order, 
and (ii) the assumption that slightly away from half-filling 
the Fermi surfaces still retain the tilted square shape that 
they had at half filling. The second assumption is not obviously 
valid at large magnetization in the attractive problem, even if 
the mean density is $n=1$, and makes the approximation unreliable. 
Nevertheless it establishes a simple benchmark at small polarization.
}

Fig.3 shows the $\mu - h$ and $n - m$ phase diagrams inferred from the
variational calculation. We have demarcated the regions pertaining to 
the different
wave vectors. There is no noticeable window of phase
separation in the $n-m$ plot. 
The number density shows small discontinuity
between the phases with different modulation
 wave vectors, as a consequence of
the finite size of the lattice.
The transitions are essentially continuous and we expect the
$q$ variation to become continuous as $L \rightarrow \infty$.

\subsection{Density of states}

Unlike the homogeneous superfluid 
the modulated state gives rise to additional
features in the energy spectrum \cite{yang2012, ting2006}.
Fig. 4 shows the spin up and total DOS as 
calculated through BdG diagonalization in 
comparison to those obtained through 
Green's function formalism.
These will serve as a reference when examining
the thermal effects.  The DOS
deviates considerably from the homogeneous 
superfluid case, and when finite size effects are
eliminated one expects a DOS with even more
intricate features \cite{yang2012, ting2006}, as shown 
through the Green's function results.
The Green's function 
shows how the effect of scattering 
from ${\bf k}_{\uparrow}$ to ${\bf -k+Q}_{\downarrow}$
and ${\bf -k-Q}_{\downarrow}$
in presence of the order parameter modulation.

{For any finite ${\bf Q}$, the DOS deviates considerably from 
that of its BCS (${\bf Q} = 0$) counterpart. There is finite 
DOS at $\omega = 0$ and there are visible new van Hove 
singularities.
These arise from the ${\bf k}$ 
regions where the dispersion $E_{\alpha}({\bf k})$ 
satisfies the condition $\partial E_{\alpha}({\bf k})/\partial {\bf k} = 0$,
where, $\alpha$ correspond to the multiple branches 
that arises in the dispersion of the LO state.
}

In Fig.4f we show the spin up spectral map for a 
$(0,0) \rightarrow (\pi,\pi)$ momentum scan
calculated through the BdG formalism, and compare the 
same with the result from the Green's function 
approach, 4e. The dispersion shows three 
distinct branches separated by `gaps' 
These regions of suppressed weight (not isotropic on the
Brillouin zone) lead to the principal depressions 
in the DOS shown in Fig.4a and 4c. 

\section{Results: thermal behavior}

\subsection{Phase diagram}

Our first paper presented the overall $h-T$ 
phase diagram at $U=4t$. We observed that 
the unpolarised 
superfluid ground state, with ${\bf q} =0$,
undergoes a first order transition to
a ${\bf q} \neq 0$ LO state at some $h = h_{c1}$ \cite{conduit}.
The modulation wave vector of this state,
and its magnetization, grows with field till, at $h= h_{c2}$,
the order is lost to a 
partially polarized Fermi liquid through a second order transition \cite{torma2014}.
For the $U$
and $\mu$ that we have chosen, $h_{c1} \sim 0.85t$
while $h_{c2} \sim 1.25t$.
$h_{c1}$ and $h_{c2}$ define the field boundary in
the phase diagram in Fig.5.

Fig.5(a)-(b) presents the $h-T$ phase
 diagram determined from our Monte
Carlo calculation. The left panel shows the LO window
estimated from the cluster based MC,
with the superposed 
dotted line indicating the $T_c$ estimated 
from a small size exact diagonalization (ED) based MC. 
We have performed this check since the energy of small
${\bf q}$ LO states is sensitive to system size and
cluster based approximations can give an erroneous
estimate of $T_c$. 
The consistency with ED based MC suggests that
our $T_c$ estimates are reasonable.

\begin{figure}[b]
\centerline{
 \includegraphics[height=5.4cm,width=4.2cm,angle=0]{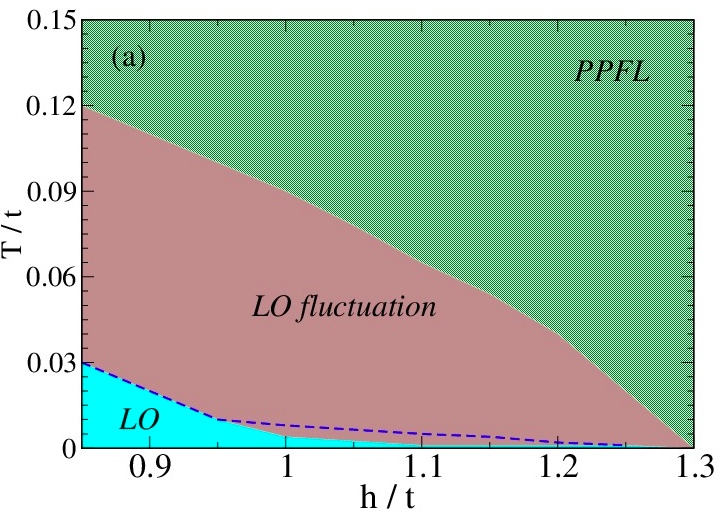}
 \includegraphics[height=5.4cm,width=4.2cm,angle=0]{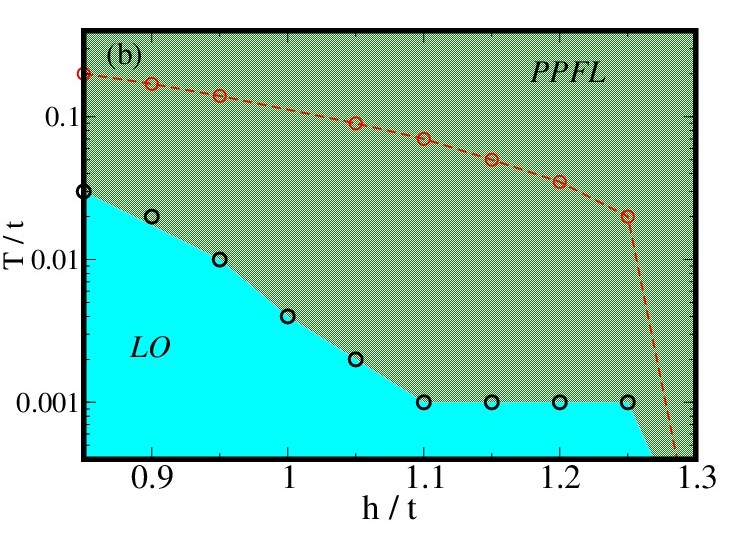}
}
\centerline{
 \includegraphics[height=3.5cm,width=4.2cm,angle=0]{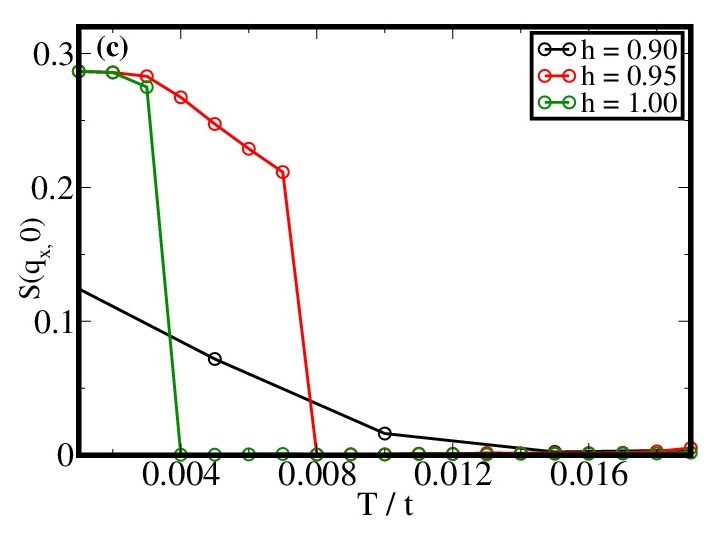}
 \includegraphics[height=3.5cm,width=4.2cm,angle=0]{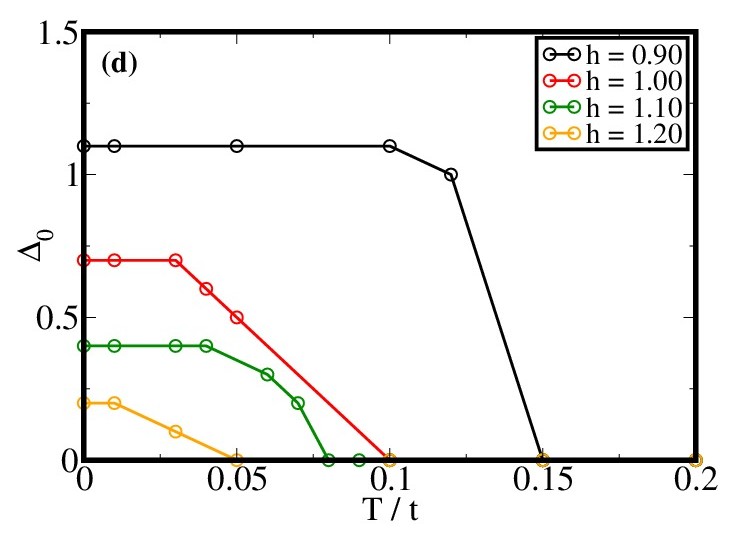}
}
 \caption{Color online: $h-T$ phase diagram and order parameter.
(a).~MC based $h-T$ phase diagram at $\mu = -0.2$ obtained
on a $24 \times 24$ lattice through heating cycle.
It indicates the LO ordered and LO fluctuation regimes (see text).
 The dashed line corresponds to the $T_{c}$ inferred from a
 $ 10 \times 10$ ED based MC calculations. (b).~$h-T$ phase diagram
comparing the MC and MFT based $T_c$ estimates. The $T$ axis is
logarithmic. (c)~The temperature dependence of the ordering peak in
$S_{\Delta}({\bf q})$ at different $h$. (d)~The temperature
dependence of the MF order parameter. The temperature range in (c)
and (d) differ by more than a factor of 10.
}
\end{figure}

\begin{figure*}
\includegraphics[height=5.5cm,width=17.5cm,angle=0]{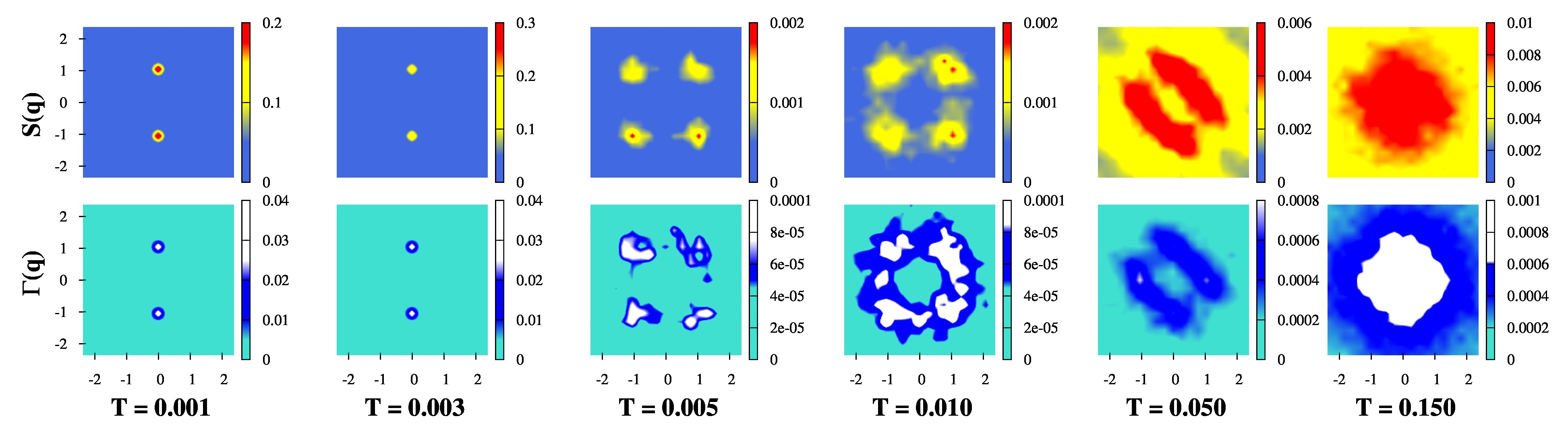}
\vspace{-.3cm}
\includegraphics[height=5.5cm,width=17.5cm,angle=0]{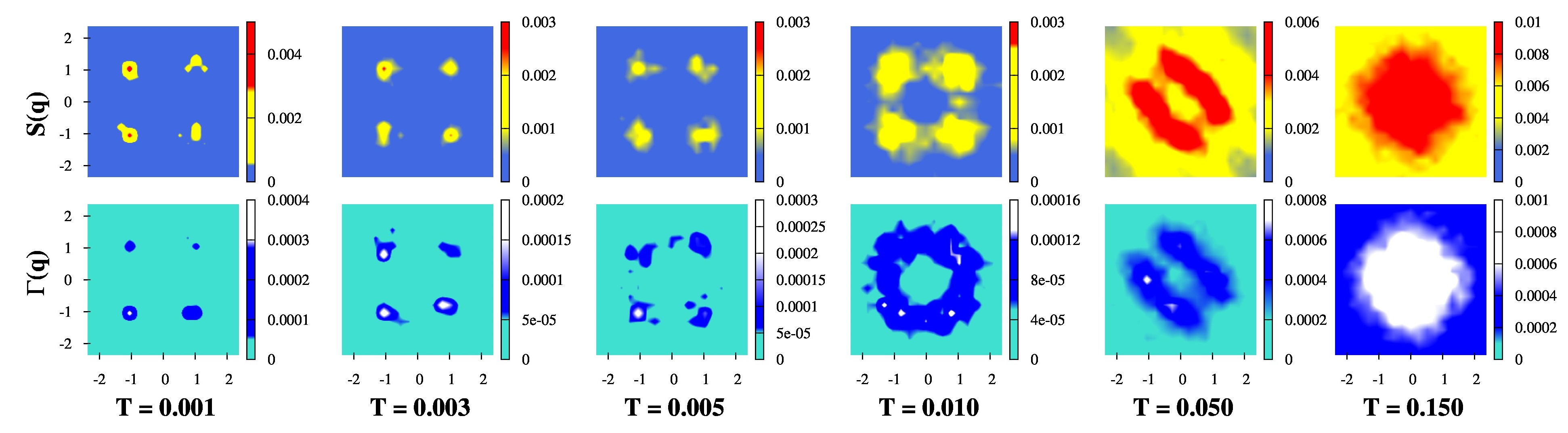}
\caption{Color online: Thermal evolution of the structure factors
associated with pairing, $S_{\Delta}({\bf q})$ and $\Gamma({\bf q})$.
The $x$ and $y$ axes in the panels represent $q_x$ and $q_y$, both
ranging from $-\pi$ to $\pi$.
The upper set of panels is for heating from the variational LO state
and the lower set for cooling from a random state.
The structure factors display a ${\bf q} \neq (0,0)$ centered
feature well above $T_c$, upto $T \sim 0.05t$.
The cooled system, at the lowest $T$ is unable to
organize itself into a perfectly ordered axial phase.
}
\end{figure*}

The typical $T_c$ scales
are $\sim 0.01t$. The wide `LO fluctuation' window 
above $T_c$ 
is characterized by the presence of finite ${\bf q}$ signature
in the pairing structure factor, $S_{\Delta}({\bf q})$,
and extends to $T \sim 0.1t$ near $h_{c1}$.
We will show the $S_{\Delta}({\bf q})$ and $\Gamma({\bf q})$ 
further
on and note that in the `balanced' superfluid such 
fluctuation would be centered at ${\bf q} =0$.

Fig.5.(b) compares the MC inferred $T_c$ to the mean 
field estimate. The MF $T_c$ differs from the MC estimate by 
a factor of $\sim 6$ at $h_{c1}$, and a factor $> 10$ 
in the middle of the LO window. The temperature axis in
Fig.5.(b) is logarithmic.
The lower two panels in Fig.5 show the MC based order
parameter (on the left) and the MFT based order parameter
(on the right). They both involve strong first order 
transitions, but the $T_c$ scale for MFT is far higher
as we have already noted.

\subsection{Structure factor}

In Fig.6 we show the thermal evolution 
of the structure factor associated with pairing.
We have explored both
heating (from the variational state) and cooling
(from an uncorrelated high $T$ state).
On heating from the variational ground state the 
axial stripes disorder (as would be more obvious
from the spatial patterns later) and the structure
factor peaks at $(0,\pm Q)$ weaken and the system
jumps discontinuously to the unordered but
weakly correlated phase. The disordered phase has a 
fourfold symmetric feature in the structure factor
for $T$ just above $T_c$  and this distorts into
a more diagonal pattern by the time $T \sim 10T_c$ and
is replaced by the usual broad feature around
${\bf q} = (0,0)$ by $T \gtrsim 0.1t$. The highest $T$
plot shows this feature. The fermionic correlation
$\Gamma({\bf q})$ roughly follows the behavior of 
$S_{\Delta}({\bf q})$.

\begin{figure}[b]
\includegraphics[width=4.0cm,height=4.8cm,angle=0]{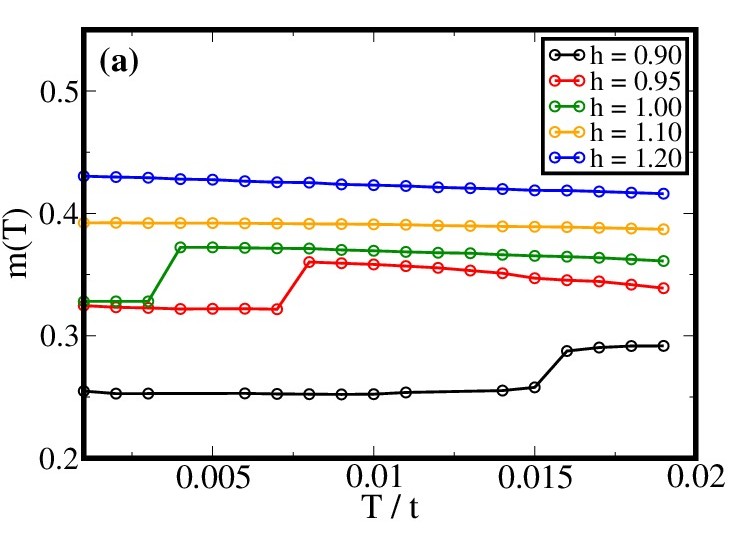}
\includegraphics[width=4.0cm,height=4.8cm,angle=0.0]{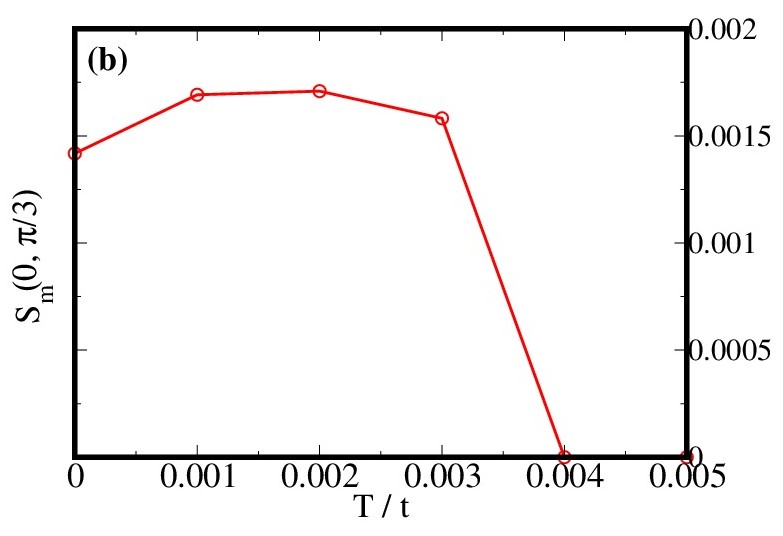}
\caption{Color online: 
Temperature dependence of (a)~magnetization for varying $h$,
and (b)~the AF peak in the magnetic structure factor at $h=1.0$}
\end{figure}

\begin{figure*}[t]
\includegraphics[height=8.5cm,width=16.0cm,angle=0]{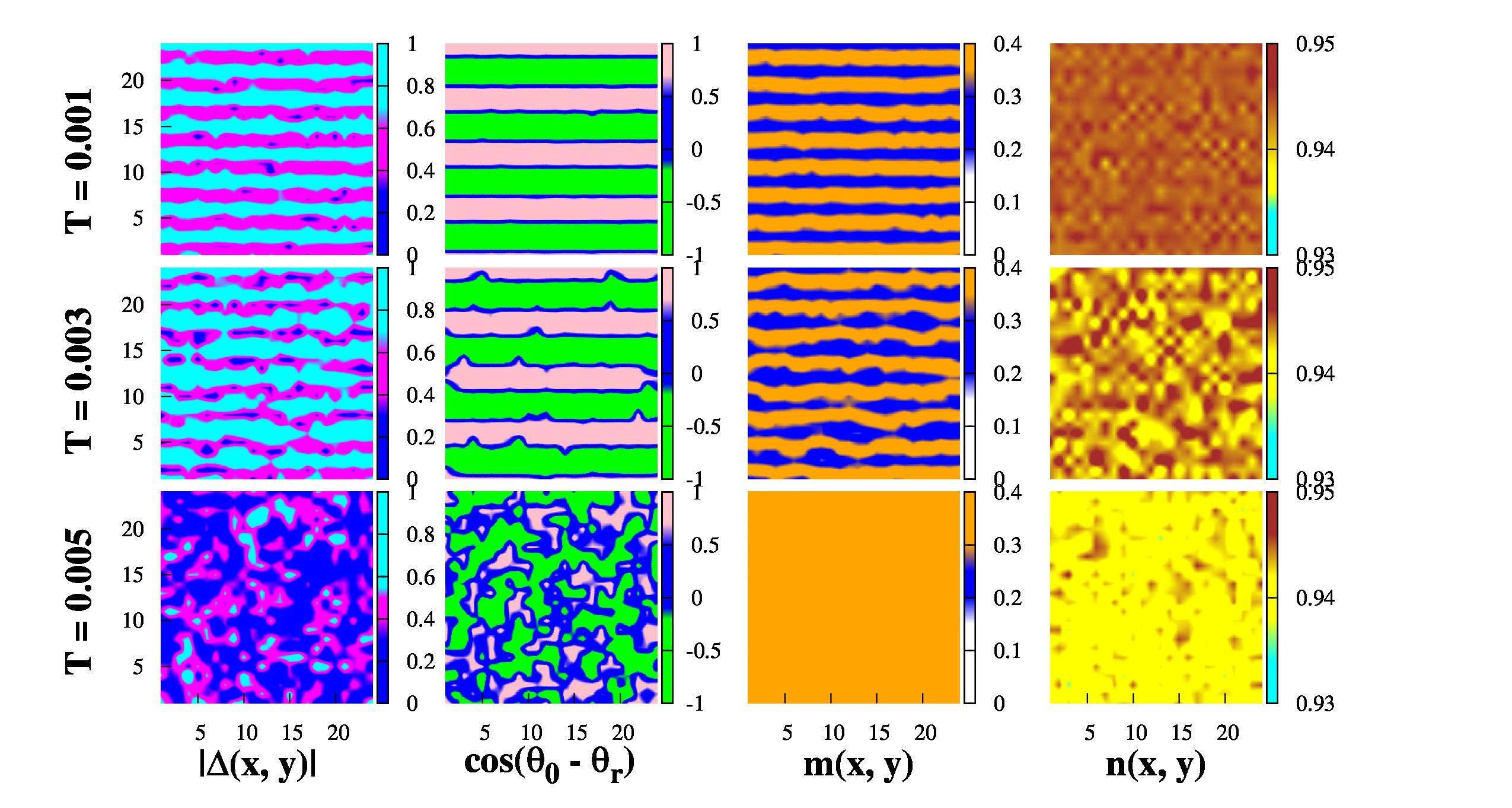}
\caption{Color online:
Spatial maps characterizing the thermal evolution of the
LO state at $h=1.0t$, through
(a)~pairing field magnitude ($\mid \Delta_{i}\mid$),
(b)~phase correlation ($\cos(\theta_0 - \theta_{i})$, $\theta_0$ is a reference site),
(c)~magnetization ($m_{i} = n_{\uparrow} - n_{\downarrow}$)
and (d)~number density ($n_{i} = n_{\uparrow} + n_{\downarrow}$).}
\end{figure*}
\begin{figure*}[t]
\includegraphics[height=7cm,width=18cm,angle=0]{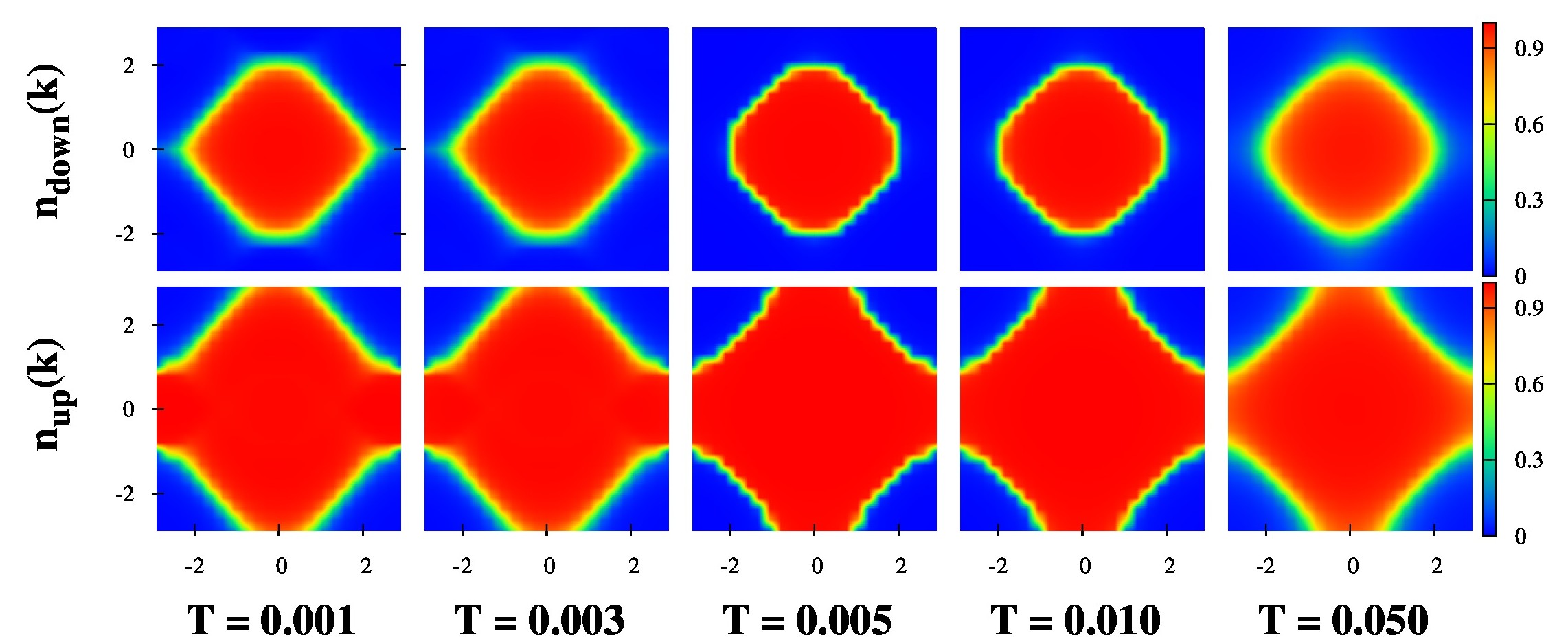}
\caption{Color online: Thermal evolution of the momentum
occupation number $n_{\sigma} ({\bf k})$ at h = 1.0t.}
\end{figure*}

On cooling the pairing structure factor retraces
the heating features down to $T \sim T_c$ but then
transits to a somewhat different, and poorly ordered,
modulated state of the form $\Delta_i \sim e^{iQx}cos(Qy)$ 
at low $T$.  This state is 
energetically higher than the variational ground
state, $\Delta_i \sim cos(Qy)$. We discuss about this 
observation below.

In the presence of metastable states the result of changing 
parameters like field or temperature can be path dependent.
This is particularly true of Monte Carlo calculations where the 
system is supposed to explore the whole energy landscape and, 
for extended time, may remain trapped in local minima. As a 
consequence the 'heating' and 'cooling' cycles might lead to 
different low temperature states as shown in Fig.6. This is 
a computational observation, within the limits of system 
size, timescales and update method used by us.

The major difference, with respect to this 
in the real system, would 
arise from two sources (i)~the much larger size leads to a 
different pattern of metastability, and (ii)~timescales:
Monte Carlo accesses about $10^{4}$ sweeps per temperature, 
and the system can remain trapped on that timescale, while 
real life systems involve timescales $\sim 10^{-9}$s, i. e. 
$\sim 10^{9}$ microscopic moves per second and may be able 
to escape metastable states on laboratory timescales. 

The associated magnetic structure factor, not shown here,
shows a prominent $\{0, 0\}$ peak,
corresponding to bulk magnetization,
and a feature at finite ${\bf q}$
arising from 
ripples in the $m_i$, induced by the 
modulation in $\Delta_i$. 
This `antiferromagnetic' peak is usually the
only direct signature of a LO state.
Neutron scattering experiments carried out on Pauli
limiting superconductors do show such
a feature \cite{kenzelmann2014}. 
Our finite size calculation makes it difficult to
establish the lineshape of the AF peak, and anyway
above $T_c$ this peak is too faint to be visible
compared to the bulk magnetization.

Fig.7(a) shows the $T$ dependence of the 
net magnetization, $m = \langle n_{i\uparrow} - 
n_{i \downarrow}\rangle$, at different magnetic fields.
At lower $h$ the system shows a 
first order transition at 
 $T = T_{c}$, and for $h \gtrsim 1.1t$ the $m(T)$ is
smooth - consistent with a very low ordering scale
(our $h_{c2}$ estimate is $\sim 1.2t$).
Fig.7(b) shows the temperature dependence 
of the antiferromagnetic peak in $S_m({\bf q})$. 
The discrete momenta on the $24 \times 24$ lattice makes it
difficult to extract a meaningful lineshape for the
AF peak.

\begin{figure*}[t]
\centerline{
\includegraphics[width=5.8cm,height=5.3cm,angle=0]{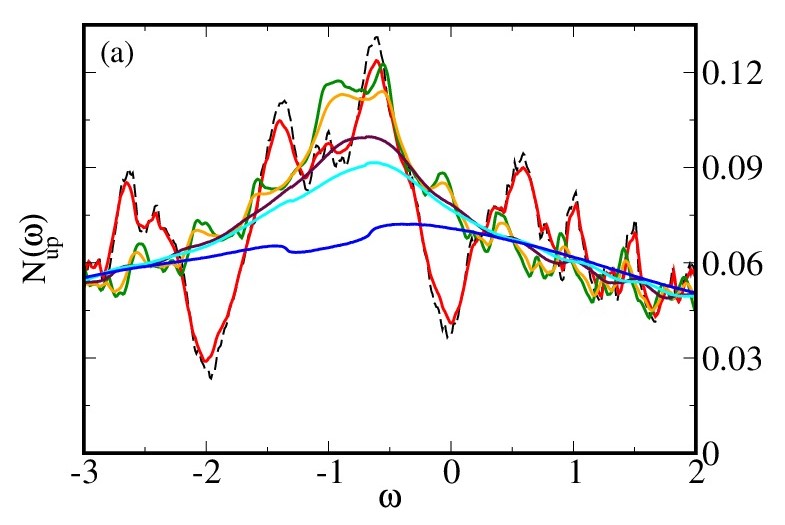}
\includegraphics[width=5.8cm,height=5.3cm,angle=0]{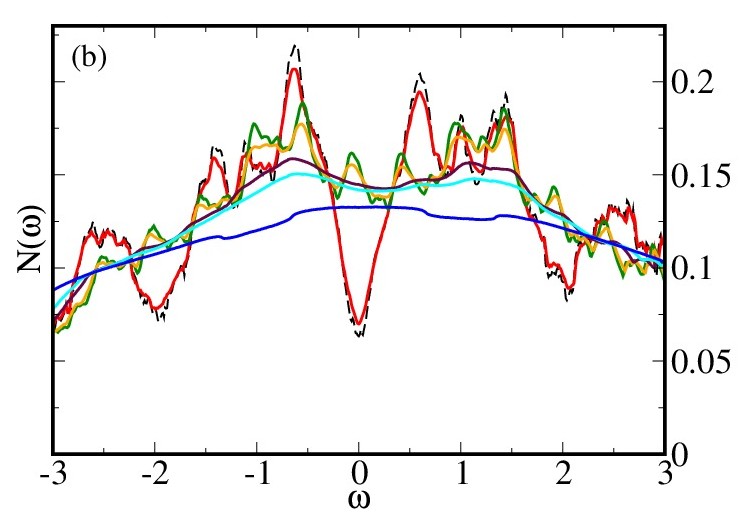}
\includegraphics[width=5.8cm,height=5.3cm,angle=0.0]{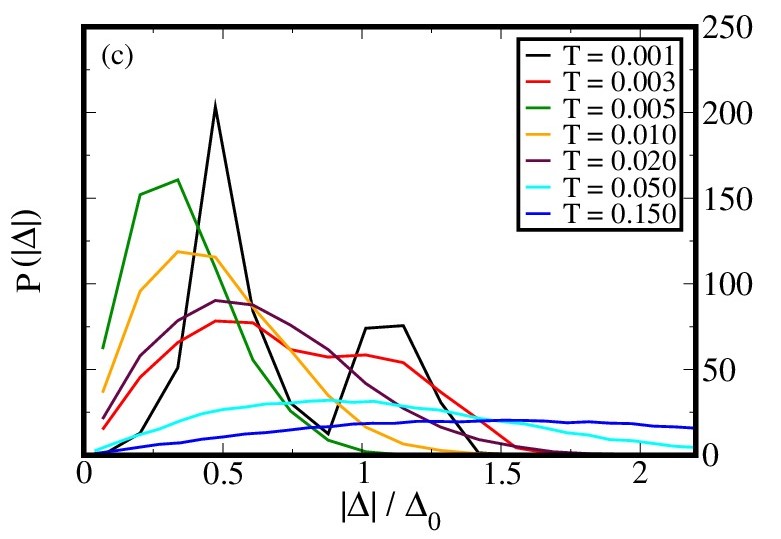}
}
\caption{Color online: Density of states and the distribution
function for $\vert \Delta \vert$. (a).~Temperature dependence
of the up spin DOS $N_{\uparrow}(\omega,T)$ at $h=t$,
(b)~Temperature dependence
of the total DOS $N(\omega,T)$ at $h=t$, and
(c)~$P(\vert \Delta \vert,T)$ at $h =t$.
The thermal evolution wipes off the delicate features of the low 
temperature DOS, associated with Andreev reflection.}
\end{figure*}

\subsection{Spatial behavior}

Apart from the structure factor, the FFLO
state can be characterized through
real space signatures viz. pairing field amplitude,
phase correlation, magnetization
and number density distribution. We present this
in Fig.8. 
The figure clearly shows the axial stripe character of the
paired state with the first two, low temperature, rows
revealing that the modulation in $\vert \Delta_i \vert$ and
$m_i$ have the same period while the phase variable has
double the wavelength. The `nodes' in 
$\vert \Delta_i \vert$
correspond roughly to the peaks in $m_i$.
The local density $n_i$ remains within $1\%$ of $0.94$.
The third row is for $T > T_c$ and there are no obvious
signatures of even short range correlation in the amplitude
and phase variables, while $m_i$ is essentially 
homogeneous. Nevertheless, as $S_{\Delta}({\bf q})$ reveals,
weak finite ${\bf q}$ correlations survive to a fairly
high temperature.

\subsection{Momentum distribution}

We compute the spin resolved fermion momentum 
distribution and show the result for the heating
cycle at $h=t$ in Fig.9.
The 
symmetry is clearly 
two fold due to the axial modulation. 
In the FFLO state the pairing is between ${\bf k},\uparrow$
and $-{\bf k}+{\bf Q}, \downarrow$, where the ${\bf k}$'s
are supposed to be on the up Fermi surface (FS)
 and the ${-\bf k} + {\bf Q}$ on the down FS.
The ${\bf Q}$ inferred from the structure factor and
the spatial maps is $ \pm {\hat y} \pi/3$ for the axial
LO state.

To check how well the paired momenta lie on the respective FS
 we varied ${\bf k}$ on the up spin FS and located the
corresponding $- {\bf k} \pm {\hat y} \pi/3$  on the down
spin FS. For ${\bf k}$ in the upper branches of the up spin FS
the $+ {\hat y} \pi/3$ connection generates points that are
reasonably close to the smaller down spin FS. The correspondence
is best for points along the diagonal and worsens as we move ${\bf k}$
along the FS to the zone boundary. The complementary behavior holds
for the lower branches of the FS when we consider the
$- {\hat y} \pi/3$ momentum transfer.

For ${\bf k}$ neighbourhoods
where {${\bf k},\uparrow$}
fails to find a partner as {${\bf -k+Q},\downarrow$} the gap is 
smaller, and a seemingly nodal gap structure, arising due 
to the LO modulation, can be seen via the 
spectral weight distribution. We have discussed about these 
spectral features of the modulated LO phase, elsewhere \cite{mpk_spec}. 

Now the thermal evolution.
For $T < T_{c} \sim 0.005t$ the 
distributions show only a two fold symmetry and of course 
differing sizes due to the magnetization in the LO state.
For $T > T_{c}$ and up to $\sim 2T_c$ the
distributions are almost fourfold symmetric but have some 
weak modulation. By the time $T \sim 0.05t \sim 10T_c$
the distributions only reflect the population imbalance of
the PPFL, with no trace of modulation.

\subsection{Density of states}

Although most of the `sightings' of a FFLO state are in
the solid state, there seem to be no detailed data
confirming its unusual spectral features.
In continuum atomic gases prominent pseudogap 
features have been observed in the balanced case
\cite{chin2005, strinati_nat_phys}
and  analyzed in detail theoretically \cite{strinati2012, 
strinati2002}. For the imbalanced gases also both 
experimental and theoretical studies suggest
pseudogap features at large imbalances
\cite{ketterle_science2007, levin2006, levin2008, levin2009}, 
even though the detailed angle resolved measurements have not
been made. Finite $T$ results on the FFLO state on a
lattice seem to be rare. We discuss the `pseudogap'
features in this regime based on our density of states
observations, below.

We focus on the 
following: (i)~The low temperature spectrum that
is characteristic of the modulated pair state,
with its multiple peaks and dips, (ii)~The thermal
evolution of the ideal $T=0$ spectrum in the $T < T_c$
window, and (iii)~The signatures of large $U$,
and possibly short range pairing correlations, 
in the $T > T_c$ window. Note that in the Hubbard
model, {\it i.e}, with contact interactions, the
effectiveness of Fermi-Fermi interactions weaken 
with increasing population imbalance - so the
survival of a high $T$ pseudogap at large population
imbalance is not obvious \cite{mpk_pg}.

Fig.10(a)-(b) shows the spin up and 
spin summed fermion density 
of states for $h = t$ corresponding to 
 ${\bf Q} = \{0, \pi/3\}$. 
The up spin spectrum in the ground state 
has a depression \cite{yang2012}  at $\omega \sim -h$,
{\it i.e}, the LO ground state 
is `pseudogapped' (PG), 
due to its peculiar band structure, unlike the gapped `BCS state' 
\cite{mpk_bp, mpk_pg}.
Increasing $T$ quickly weakens these features,
and the strong first order transition at $T_c$ leads
essentially to a gapless state. 
$T_{c}$ therefore is also the PG to `ungapped' transition
point for the spin resolved DOS. 
The gapless phase survives to a scale 
$T_{pg2}$ \cite{mpk_bp} beyond which the 
system {\it again shows a weak pseudogap}. This `re-entrant'
feature is due to the thermally induced 
large amplitude fluctuations in this strongly interacting,
$U=4t$, problem \cite{mpk_bp, mpk_pg}.

The observed spectral features can be analyzed 
based on the 
distribution $P(\vert \Delta \vert)$.
In Fig.10(c) we have shown
the distribution of the pairing field at temperatures corresponding
to those of the DOS. For $T \le T_c$ $P(\vert \Delta \vert)$ is
multi-peaked due to the amplitude modulated LO state.
Increasing $T$ beyond $T_c$ makes the distribution broad, single 
peaked, and reduces the mean magnitude of $\Delta_i$.
The weak pairing field leaves the DOS essentially featureless.
For $T > 0.05t$ the behavior
of the distribution changes significantly. The mean of
$P(\vert \Delta \vert)$ shifts to significantly high values
with increasing $T$, although the distribution remains
very broad. 
This, we believe, is similar to the PG effect seen at
large imbalance in the unitary Fermi gas.

\section{Discussion}

We have discussed the strength and limitations of the static auxiliary
field scheme in Paper I \cite{mpk_bp}. 
There are size limitations that are specific to the FFLO
phase which we touch upon here. We also discuss experimental situations
where our results can shed some light.

\subsection{Methodological issues}

\subsubsection{Single channel decomposition}

We have decomposed the Hubbard interaction only
in the pairing channel. 
At $T=0$ our 
auxiliary field approach, using the field $\Delta_{i}$, reduces 
to a 'pairing channel' mean field theory, the standard BdG scheme.
Mean field theory can be implemented in more complex formats, 
by decomposition in the pairing, density, and spin channels 
simultaneously. This is the Hartree-Fock BdG (HFBdG) approach 
and is obviously more general than simply BdG. Such calculations
yield diagonal stripes over a part of the phase diagram 
\cite{zhang2013} while we obtain only axial stripes.

Unfortunately, the multichannel mean field theories are not 
good templates for including fluctuations beyond the Gaussian 
level. Since the strength of our method is really in calculating 
the thermal properties we opted to use the single channel 
Hubbard-Stratonovich so that large amplitude fluctuations can be 
systematically included, and the normal state well captured.

\subsubsection{Quantum fluctuations in the FFLO state}


We use a static auxiliary field 
technique where the temporal (quantum) fluctuations 
of the pairing field are neglected, while 
spatial fluctuations are completely retained.
While this could be a poor approximation in
the continuum, on a lattice it is reasonable as
we argue below.

The low energy 
fluctuations in the continuum FFLO state, {\it i.e}, in free space,
arise from three sources (i)~the `phase symmetry' of the $U(1)$ order
parameter, (ii)~the translational symmetry breaking, and (iii)~the
rotational symmetry breaking \cite{radz2011}. As a result, in
two dimensions, long range order cannot be sustained even at $T=0$
and mean field theory (which predicts such order) is invalid.

On a lattice the phase field 
still has low energy excitations of the `XY' type, but the translational
and rotational modes would be gapped out due to the spatial symmetry already
broken by the underlying lattice \cite{loh}. 
Models with XY symmetry have long range order in the 2D ground state and
a KT transition at finite $T$. Therefore, the issue of fluctuations
reduces to checking how well the $U(1)$ superfluid $T_c$ is captured by
our model, in 2D, {\it vis-a-vis} full QMC. A comparison between the 
QMC results and that obtained by the present numerical technique has been 
made in ref.\cite{tarat2014}. However, for a spin imbalanced system 
such QMC results do not seem to exist. Nevertheless, the argument
above and the $T_c$ comparison for the balanced system shows that, 
for the fluctuations that are relevant, our method does quite well.

\subsubsection{Finite size effects in the FFLO phase:-}

The variational ground state calculation for the 
FFLO state is affected by the finite size of the 
lattice. The number of ${\bf q}$ values (modulation 
wave vector of the LO state) is determined by the 
size of the lattice.

A more serious size limitation arises for the finite 
temperature Monte Carlo simulation owing to the cluster
update technique that we use \cite{mpk_bp}. Even though 
the cluster update technique allows us to access large 
system sizes ($\sim 30 \times 30$) within reasonable 
computation time, it introduces another length scale 
in the problem. Consequently, within our approach 
there are two sources of error.  (i)~The finite wavelength,
$\lambda$, of the FFLO modulations require the linear dimension
of the system, $L$ to be much greater than $\lambda$.  (ii)~If
the energy cost of MC update is not computed via exact 
diagonalization of the $L \times L$ system, but on a $L_c \times L_c$ 
cluster, one needs $L_c \gg \lambda$.
These are difficult constraints to satisfy when trying to access
large $\lambda$, {\it i.e}, small $Q$ states.
As a result most of our detailed thermal results are on
relatively large ${\bf Q}$ states.

\subsection{Connection to experiments:-}

As discussed in the introduction the FFLO state has
so far been realized experimentally only in the solid state,
with some evidence in heavy fermions \cite{bianchi2003, 
kenzelmann2014, tayama2002, mitrovic2008, kumagai2006, mitrovic2006,
wright2011, kenzelmann2008}, organics \cite{lortz2007, beyer2013, coniglio2011,
mayaffre2014, bergk2011, agosta2012, cho2009} and
iron pnictides \cite{zocco2014, prozorov2011, kim2011}.
However, most of
the superconductors are non $s$-wave, 
and are at considerably weaker coupling than we have considered. 
None of them are close to their BCS-BEC crossover coupling, so 
a detailed comparison of our thermal results with them is
inappropriate.

Turning to cold atoms, the key signature of a LO state is the 
spatial modulation of the pairing field and polarization.
Moreover, the LO state has finite polarization
down to zero temperature, unlike the breached pair state where
the polarization vanishes as $T \rightarrow 0$. 
While there are 
so far no experimental
signatures of real space polarization modulation,
the spin resolved
density profiles in quasi one dimensional traps indicate 
finite polarization down to 
the lowest accessible temperature \cite{liao2010}.
The spectral features of this state has not been
measured.

\section{Conclusion}

We have studied the attractive Hubbard model
in the coupling regime of BCS-BEC crossover and
large population imbalance. 
We established the thermal properties of 
the FFLO phase that occurs in this regime.
The thermal transition 
is strongly first order
with a $T_c$ much below the
mean field prediction. 
Finite momentum pairing correlations nevertheless
persist to a scale that is $\sim 10 T_c$. We have
studied the fermionic density of states and
discover thermally induced closure of the
characteristic FFLO dip at $T \sim T_c$
but the reappearance of a correlation
induced pseudogap at higher temperature.
The $T_c$ estimate and the detailed physical
indicators should help in pinpointing FFLO states
in strong coupling cold atomic optical lattices.

{\it Acknowledgments:}
We acknowledge discussions with J. K. Bhattacharjee and
use of the High Performance Computing Cluster at HRI.
PM acknowledges support from an
Outstanding Research Investigator Award of the -SRC.



\begin{thebibliography}{99}
\bibitem{ff}
P. Fulde and A. Ferrell, Phys. Rev. {\bf 135},  (1964).
\bibitem{lo}
A. Larkin and Y. N. Ovchinnikov, . Eksp. . Fiz. {\bf 47},
1136 (1964) [Sov. Phys. JETP {\bf 20}, 762 (1965)].
\bibitem{bianchi2003}
A. Bianchi, R. Movshovich, C. Capan, P. G. Pagliuso and J. L. Sarrao, Phys. Rev. Lett.
{\bf 91}, 187004 (2003).
\bibitem{kenzelmann2014}
S. Gerber, M. Bartkowiak, J. L. Gavilano, E. Ressouche,
N. Egetenmeyer, C. Niedermayer, A. D. Bianchi,
R. Movshovich, E. D. Bauer, J. D. Thompson and M. Kenzelmann, 
Nat. Phys. {\bf 10}, 126-129 (2014).
\bibitem{tayama2002}
T. Tayama, A. Harita, T. Sakakibara, Y. Haga, H. Shishido, R. Settai and Y. Onuki,
Phys. Rev. B {\bf 65}, 180504(R) (2002).
\bibitem{mitrovic2008}
G. Koutroulakis, V. F. Mitrovic, M. Horvatic, C. Berthier,
G. Lapertot and J. Flouquet, Phys. Rev. Lett. {\bf 101}, 047004 (2008).
\bibitem{kumagai2006}
K. Kumagai, M. Saitoh, T. Oyaizu, Y. Furukawa, S. Takashima, M. Nohara, H. Takagi
and Y. Matsuda, Phys. Rev. Lett. {\bf 97}, 227002 (2006).
\bibitem{mitrovic2006}
V. F. Mitrovic, M. Horvatic, C. Berthier, G. Knebel, G. Lapertot and 
J. Flouquet, Phys. Rev. Lett. {\bf 97}, 117002 (2006).
\bibitem{wright2011}
J. A. Wright, E. Green, P. Kuhns, A. Reyes, J. Brooks, J. Schlueter,
R. Kato, H. Yamamoto, M. Kobayashi and S. E. Brown, Phys. Rev. Lett. {\bf 107}, 
087002 (2011).
\bibitem{kenzelmann2008}
M. Kenzelmann, Th. Strassle, C. Niedermayer, M. Sigrist, B. Padmanabhan, M. Zolliker,
A. D. Bianchi, R. Movshovich, E. D. Bauer, J. L. Sarrao and J. D. Thompson, Science {\bf 321},
1652 (2008).
\bibitem{lortz2007}
R. Lortz, Y. Wang, A. Demuer, P. H. M. Bottger, B. Bergk, G. Zwicknagl,
Y. Nakazawa and J. Wosnitza, Phys. Rev. Lett. {\bf 99}, 187002 (2007).
\bibitem{beyer2013}
R. Beyer and J. Wosnitza, Low Temp. Phys. {\bf 39}, 225 (2013).
\bibitem{coniglio2011}
W. A. Coniglio, L. E. Winter, K. Cho and C. C. Agosta, 
B. Fravel and L. K. Montgomery, Phys. Rev. B {\bf 83} 224507 (2011).
\bibitem{mayaffre2014}
H. Mayaffre, S. Kr̈amer, M. Horvati ́c, C. Berthier, K. Miyagawa,
K. Kanoda V. F. Mitrovic, Nat. Phys. {\bf 10}, 928 (2014).
\bibitem{bergk2011}
B. Bergk, A. Demuer, I. Sheikin, Y. Wang, J. Wosnitza,
Y. Nakazawa and R. Lortz, Phys. Rev. B {\bf 83}, 064506 (2011).
\bibitem{agosta2012}
C. C. Agosta, Jing Jin, W. A. Coniglio, B. E. Smith,
K. Cho, I. Stroe, C. Martin, S. W. Tozer, T. P. Murphy, E. C. Palm,
J. A. Schlueter and M. Kurmoo, Phys. Rev. B {\bf 85}, 214514 (2012).
\bibitem{cho2009}
K. Cho, B. E. Smith, W. A. Coniglio, L. E. Winter, C. C. Agosta
and J. A. Schlueter, Phys. Rev. B {\bf 79}, 220507(R) (2009).
\bibitem{zocco2014}
D. A. Zocco, K. Grube, F. Eilers, T. Wolf and H. v. Lo ̈hneysen, 
Phys. Rev. Lett. {\bf 111}, 057007 (2013).
\bibitem{prozorov2011}
K. Cho, H. Kim, M. A. Tanatar, Y. J. Song, Y. S. Kwon, W. A.
Coniglio, C. C. Agosta, A. Gurevich and R. Prozorov, Phys. Rev.
B 83, 060502(R) (2011).
\bibitem{kim2011}
S. Khim, B. Lee, J. W. Kim, E. S. Choi, G. R. Stewart and K. H.
Kim, Phys. Rev. B 84, 104502 (2011).
\bibitem{ptok2013}
A. Ptok and D. Crivelli, J. Low. Temp. Phys. 172, 226 (2013).
\bibitem{ptok2014}
A. Ptok, Eur. Phys. J. B 87, 2 (2014).
\bibitem{partridge2006}
G. B. Partridge, Wenhui Li, Y. A. Liao, R. G. Hulet, M. Haque
and H. T. C. Stoof, Phys. Rev. Lett. 97, 190407 (2006).
\bibitem{shin2008}
Y. Shin, C. H. Schunck, A. Schirotzek and W. Ketterle, Nature
(London) {\bf 451}, 689 (2008).
\bibitem{ketterle_science2007}
C. H. Schunck, Y. Shin, A. Schirotzek, M. W. Zwierlein and W. Ketterle,
Science {\bf 316}, 867 (2007).
\bibitem{shin2006}
Y. Shin, M. W. Zweierlein, C. H. Schunck, A. Schirotzek and W. Ketterle,
Phys. Rev. Lett. {\bf 97}, 030401 (2006)
\bibitem{liao2010}
Y-an Liao, A. S. C. Rittner, T. Paprotta, W. Li, G. B. Partridge, R. G. Hulet,
S. K. Baur and E. J. Mueller, Nature {\bf 467}, 567 (2010). 
\bibitem{ketterle_science2008}
Yong-il Shin, Christian H. Schunck, Andre ́ Schirotzek and Wolfgang Ketterle,
Nature {\bf 451}, 689 (2008).
\bibitem{trivedi}
Y. L. Loh and N. Trivedi, Phys. Rev. Lett. {\bf 104}, 165302 (2010).
\bibitem{torma2007}
T. K. Koponen, T. Paananen, J. -P. Martikainen and P. Torma, Phys. Rev. Lett. {\bf 99}, 120403 (2007).
\bibitem{scaletter2012}
M. J. Wolak, B. Gremaud, R. T. Scalettar and G. G. Batrouni, Phys. Rev. A {\bf 86},
 023630 (2012).
\bibitem{scaletter2008}
G. G. Batrouni, M. H. Huntley, V. G. Rousseau and R. T. Scalettar, Phys. Rev. Lett. 
{\bf 100}, 116405 (2008).
\bibitem{casula2008}
M. Casula, D. M. Ceperley and E. J. Mueller, Phys. Rev. A {\bf 78}, 033607 (2008).
\bibitem{zhang2013}
S. Chiesa and S. Zhang, Phys. Rev. A {\bf 88}, 043624 (2013).
\bibitem{beck}
A. Sewer, X. Zotos and H. Beck, Phys. Rev. B {\bf 66}, 140504(R) (2002).
\bibitem{torma2012}
D. -H. Kim and P. Torma, Phys. Rev. B {\bf 85}, 180508(R) (2012).
\bibitem{torma2013}
M. O. J. Heikkinen, D.-H. Kim and P. Torma, Phys. Rev. B {\bf 87}, 224513 (2013).
\bibitem{torma2014}
M. O. J. Heikkinen, D.-H. Kim, M. Troyer and P. Torma, Phys. Rev. Lett. {\bf 113}, 
185301 (2014).
\bibitem{mpk_bp}
M. Karmakar and P. Majumdar, arxiv:1508.00393 (2015). 
\bibitem{evenson}
W. E.Evenson, J. R. Schrieffer and S. Q. Wang, J. Appl. Phys. 
{\bf 41}, 1199 (1970).
\bibitem{tarat2014}
S. Tarat and P. Majumdar, Europhys. Lett. {\bf 105}, 67002 (2014).
\bibitem{zhang2011}
J. Xu, C-C. Chang, E. J. Walter and S. Zhang, J. Phys. Cond. Mat. 
{\bf 23}, 505601 (2011).
\bibitem{yang2012}
Q. Cui, C. -R. Hu, J. Y. T. Wei and K. Yang, Phys. Rev. B {\bf 85}, 014503 (2012).
\bibitem{ting2006}
Q. Wang, H.-Y. Chen, C.-R. Hu and C. S. Ting, Phys. Rev. Lett. {\bf 96}, 
117006 (2006).
\bibitem{conduit}
G. J. Conduit, P. H. Conlon and B. D. Simons, Phys. Rev. A 
{\bf 77}, 053617 (2008).
\bibitem{mpk_spec}
M. Karmakar and P. Majumdar, in preparation.
\bibitem{chin2005}
C. Chin, M. Bartenstein, A. Altmeyer, S. Riedl, S. Jochim,
J. Hecker Denschlag, R. Grimm, Science {\bf 305}, 1128 (2004).
\bibitem{strinati_nat_phys}
J. P. Gaebler, J. T. Stewart, T. E. Drake, D. S. Jin, A. Perali, 
P. Pieri and G. C. Strinati, Nat. Phys. {\bf 6}, 569 (2010).
\bibitem{strinati2012}
F. Palestini, A. Perali, P. Pieri, and G. C. Strinati, Phys. Rev. B
{\bf 85}, 024517 (2012).
\bibitem{strinati2002}
A. Perali, P. Pieri, G. C. Strinati and C. Castellani, Phys. Rev. B
{\bf 66}, 024510 (2002).
\bibitem{levin2006}
C.-C. Chien, Q. Chen, Y. He and K. Levin, Phys. Rev. Lett. {\bf 97}, 
090402 (2006).
\bibitem{levin2008}
Y. He, C.-C. Chien, Q. Chen and K. Levin, Phys. Rev. A {\bf 77}, 
011602(R) (2008).
\bibitem{levin2009}
H. Guo, C.-C. Chien, Q. Chen, Y. He and K. Levin, Phys. Rev. A 
{\bf 80}, 011601(R) (2009).
\bibitem{mpk_pg}
M. Karmakar and P. Majumdar, arxiv:1508:00398 (2015).
\bibitem{radz2011}
L. Radzihovsky, Phys. Rev. A {\bf 84}, 023611 (2011).
\bibitem{loh}
Y. L. Loh and N. Trivedi, Phys. Rev. Lett. {\bf 104}, 165302 (2010).
\end{thebibliography}
\end{document}